\documentclass{iau}

\usepackage{amsmath}
\usepackage{graphicx}
\usepackage{multirow}

\def\apj{{ApJ}}

\def\aap{{A\&A}}   
\def\mnras{{MNRAS}}

\newcommand{\solphys}{{\it Solar Phys.}}

\begin{document}

\lefttitle{A.\ R.\ Choudhuri \& B.
 K.\ Jha}
\righttitle{Theoretical model of NSSL}

\jnlPage{1}{7}
\jnlDoiYr{2021}
\doival{10.1017/xxxxx}

\aopheadtitle{Proceedings IAU Symposium}
\editors{A. Getling \&  L. Kitchatinov, eds.}

\title{The near-surface shear layer (NSSL)
of the Sun: a theoretical model}

\author{Arnab Rai Choudhuri$^1$ and Bibhuti Kumar Jha$^2$}
\affiliation{$^1$Department of Physics, Indian
Institute of Science, Bangalore 560012, India\\ email: \email{arnab@iisc.ac.in}}
\affiliation{$^2$ Southwest Research Institute, Boulder CO 80302, USA\\ email: \email{maitraibibhu@gmail.com}}


\begin{abstract}
We present a theoretical model of the near-surface shear layer (NSSL)
of the Sun.  Convection cells deeper down are affected by the Sun's
rotation, but this is not the case in a layer just below the solar
surface due to the smallness of the convection cells there.  Based on
this idea, we show that the thermal wind balance equation (the basic
equation in the theory of the meridional circulation which holds
inside the convection zone) can be solved
to obtain the structure of the NSSL, matching observational data 
remarkably well.
\end{abstract}


\maketitle

\section{Introduction}

There is a layer just below the solar surface of thickness
$\approx 0.05 R_{\odot}$ within which the angular velocity of the Sun
decreases with radius by a few percent.  This near-surface
shear layer (NSSL) is clearly visible in any map of solar internal
differential rotation produced by helioseismology: see Figure~3 of \citet{Choudhuri2021a}.  There is not yet a consensus on what gives
rise to this layer, although several different ideas have been
proposed  \citep{Guerrero2013, Hotta2015, Matilsky2019}. Here we summarize our work reported by \citet{Choudhuri2021}  and \citet{Jha2021}.

The theoretical ideas of how the two large-scale flow patterns
of the Sun---the differential rotation and the meridional 
circulation---are produced are inter-related.  A basic equation
in the theory of the meridional circulation is the thermal wind
balance equation, which holds within the body of the solar
convection zone. It is often supposed that this equation breaks
down near the solar surface surface, giving rise to a boundary
layer in the form of NSSL.  We propose the opposite view that
this equation holds even near the surface and the change in 
the nature of convection due to the small scale height
at the top layer causes the NSSL, of which the
structure can be obtained by applying the thermal wind balance
equation.

In \S2 we summarize the basic theoretical ideas about the
thermal wind balance equation and its implications for the
changed nature of convection in the top layer within the solar
convection zone.  After presenting our results in \S3, we end with our
concluding remarks in \S4.

\section{Basic theoretical ideas}

\def\pa{\partial}

We first write down the thermal wind balance equation before
explaining its significance.  One standard form of this is
\begin{equation}
 r \sin \theta \frac{\pa}{\pa z} \Omega^2 =
\frac{g}{\gamma T} \left( \frac{\pa}{\pa \theta} \Delta T \right),
\label{eq1} 
\end{equation}
where $z$ is the distance from the equatorial plane and the other
symbols have their usual meanings, $\Delta T$ being the temperature
variation with latitude to be discussed in more detail later.

Let us give an idea how the two sides of (\ref{eq1}) arise.  Since convective
motions are influenced by the Coriolis force, the nature of convection
changes slightly with latitude. The effect of the Coriolis force
on radially moving fluid blobs being the least near the poles,
the poles of the Sun are expected to be slightly hotter than
the equatorial region. This would tend to drive a flow from a pole
to the equator near the surface, i.e\ a clockwise meridional
circulation in the northern hemisphere---which is opposite of what
is seen. This effect is captured in the thermal wind term in the RHS
of (\ref{eq1}).  On the other hand, if the centrifugal force varies along
a straight line parallel to the rotation axis, that also can drive
a meridional circulation which is anti-clockwise in northern 
hemisphere for solar-like angular velocity distribution.  The
centrifugal term in the LHS of (\ref{eq1}) corresponds to this effect. The
full dynamical equation of the meridional circulation has other
terms \citep{Choudhuri2021a}.  However, a straightforward order-of-magnitude
estimate suggests that the centrifugal term is much larger
than these terms and can only be balanced by the thermal
wind term, leading to (\ref{eq1}).

Convective motions are influenced by the Coriolis force only if 
the convective turnover time is comparable to or more than the
rotation period $\approx 27$ days. This is certainly not true for
granules near the surface having turnover time of the order of a
few minutes. We thus expect a layer at the top of the convection
zone which is not affected by the Coriolis force. On the other hand,
we believe that convective motions deeper down with much longer
turnover times are affected by the Coriolis force.  Although this is
a gradual transition, we make the simplifying assumption that the
Coriolis force is important when $r <r_c$ and negligible when
$r > r_c$, where $r_c$ is a critical radius.

Now we write
\begin{equation}
    T(r, \theta) = T(r, 0) + \Delta T (r, \theta),
    \label{eq2}
\end{equation}
where $T(r, 0)$ is the radial distribution of temperature as
given by the standard solar model and $\Delta T (r, \theta)$
is the departure from it.  Since convective motions in the top
layer $r > r_c$ are not affected by rotation, we expect the
temperature gradient there to be independent of latitude, i.e.
\begin{equation}
\frac{d}{d r} T (r, \theta) = \frac{d}{d r} T (r, 0).
\label{eq3}
\end{equation}
On substituting (\ref{eq2}) in (\ref{eq3}), we arrive at
\begin{equation}
    \frac{d}{d r} \Delta T (r, \theta) = 0. \label{eq4}
\end{equation}
This suggests that $\Delta T(r, \theta)$ does not vary with
$r$ in this top layer $r > r_c$ and we can write
\begin{equation}
    \Delta T (r, \theta) = \Delta T (r_c, \theta). \label{eq5}
\end{equation}
In other words, the latitudinal variation of temperature at
the critical radius $r_c$ gets mapped to the visible surface.

We now look at the RHS of (\ref{eq1}). In the top layer $r > r_c$, the
temperature $T$ drops sharply with increasing $r$ as we
approach the surface, but (\ref{eq4}) ensures
that other quantities appearing in the RHS of (\ref{eq1}) do not vary much
with $r$ within this top layer.  This implies that the RHS of (\ref{eq1})
has to become very large close to the solar surface
due to the drop in $T$.  If (\ref{eq1}) has
to be satisfied, then the RHS of (\ref{eq1}) also has to become very large
just below the solar surface.  For this to happen, $\Omega^2$ has
to vary with $z$ rapidly.  We suggest that this is what gives rise
to the NSSL. This basic idea was suggested by \citet{Choudhuri2021}
based on order-of-magnitude estimates.  Detailed calculations were
presented by \citet{Jha2021}.

\section{Results}

\begin{figure}
    \centering
\includegraphics[width= \textwidth]{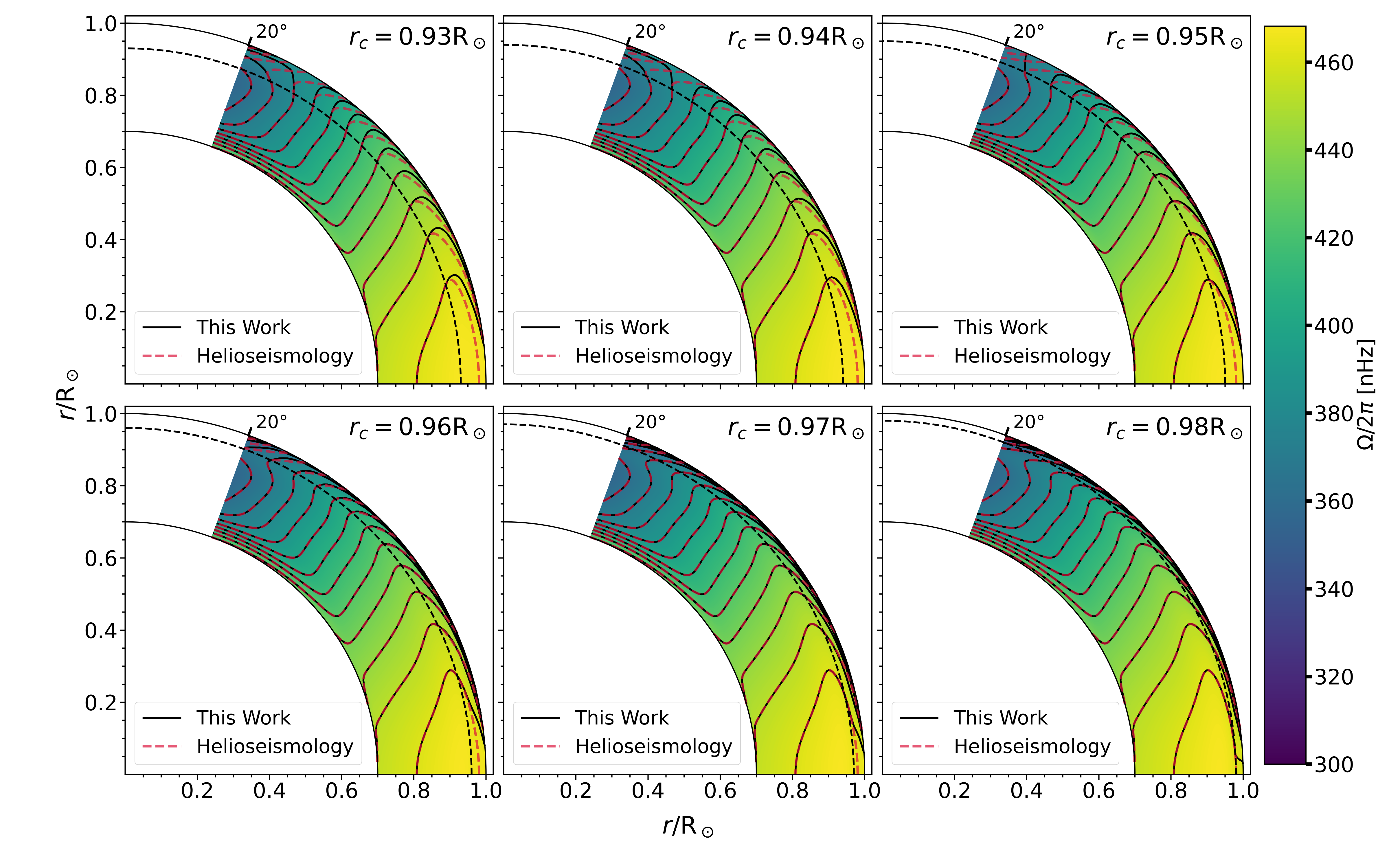}
    \caption{Contours of angular velocity $\Omega$ as given by
    helioseismolgy (dashed red curves) and by our model (solid
    black curves), for different assumed values of $r_c$. The 
    different values of $r_c$ given in different panels are
    indicated by dashed black circles. From \citet{Jha2021}.}
\label{fig1}
\end{figure}

\begin{figure}
    \centering
\includegraphics[width= 0.7\textwidth]{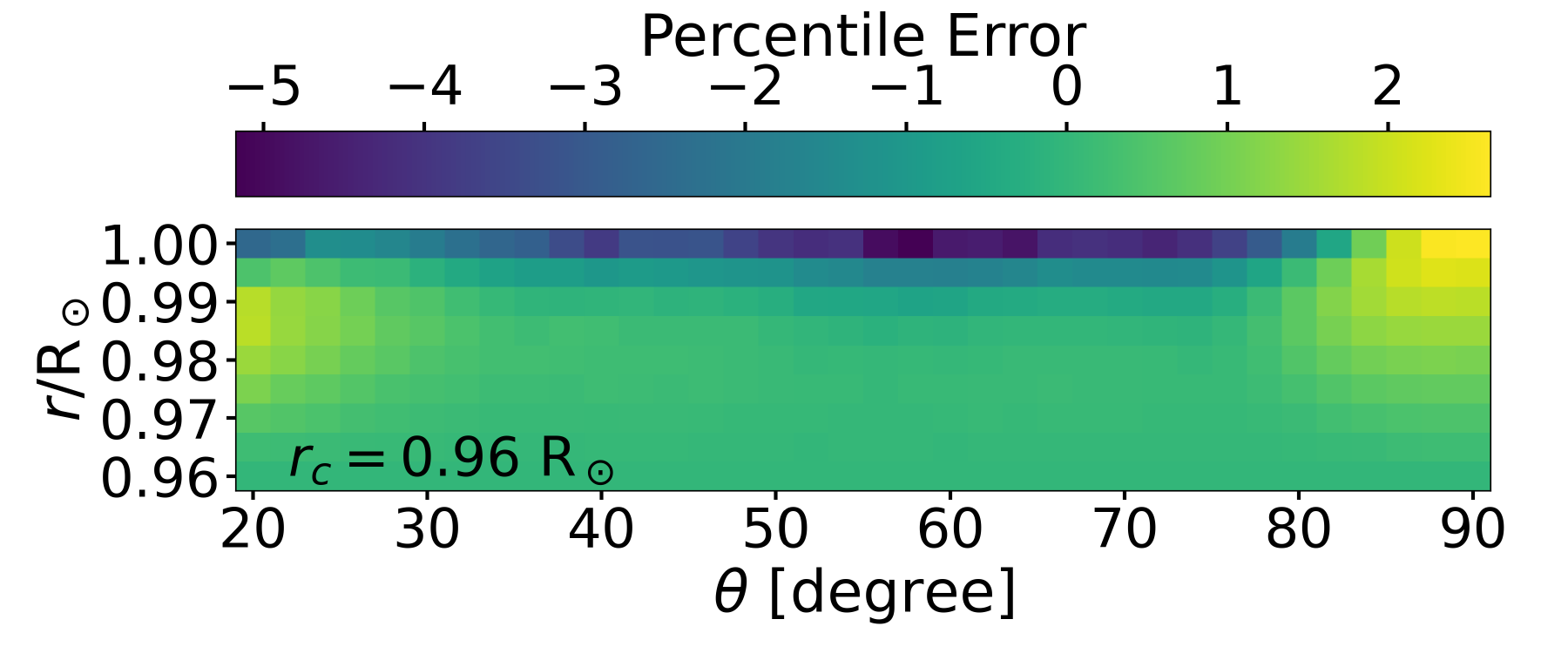}
    \caption{The percentile error of the theoretical model in
    matching the observational data of $\Omega$ for the case
    $r_c= 0.96 R_{\odot}$. From \citet{Jha2021}.}
\label{fig2}
\end{figure}

We calculate the structure of the NSSL in the top layer of
the convection zone in the following manner.
Assuming a particular value of $r_c$, we take $\Omega$ determined
from helioseismology for $r < r_c$ and find $\Delta T$ in those
deeper layers by using (\ref{eq1}). This gives us $\Delta T (r_c, \theta)$
and we easily get $\Delta T (r, \theta)$ in the top layer $r>r_c$ from (\ref{eq5}). Once we have $\Delta T (r, \theta)$ in the top layer $r>r_c$, we
can use (\ref{eq1}) to obtain $\Omega$ there, giving us the structure of
NSSL. Figure~\ref{fig1} shows different contours of $\Omega$ we get by using
different reasonable values of $r_c$.  The solid black and dashed red lines
show contours of $\Omega$ from our model and from helioseismology.
They are the same below $r_c $ because we started our calculations in
our model by using $\Omega$ from helioseismology below $r_c$. In 
the layer above $r_c$, we find that the two different sets of curves
are very close to each other and our theoretical model gives the
NSSL very nicely.  We get best results when we assume $r_c= 0.96 R_{\odot}.$ Figure~\ref{fig2} shows the percentile error in our theoretical
model in matching helioseismology data in the top layer
for $r_c= 0.96 R_{\odot}.$ Except at the very surface, we 
find the error to be much less than $5\%$.  This fantastic fit between theory
and observations gives us confidence that we are on the right track.

A side result of our theoretical model is that we get a temperature
variation at the solar surface.  For the case $r_c= 0.96 R_{\odot}$,
the pole is found to be about 3 K hotter than the equator.  It may
be noted that \citet{Kitchatinov1995} found a temperature 
difference of 5\,K in their theoretical model of large-scale flows
in the solar convection zone.  There are some observational studies \citep{Kuhn1988, Rast2008} suggesting that poles of the Sun
are indeed slightly hotter than the equator.  We hope that a more 
accurate observational study with modern techniques will be carried
out in near future.

\section{Conclusion}

For the fist time, we are able to provide a theoretical model of
the NSSL which matches observational data in quantitative detail.
This model is based on the idea that the thermal wind balance equation
holds even in the top layers of the convection zone just below the
solar surface. We get the best results on assuming that
$r_c= 0.96 R_{\odot}$ is the critical radius above which convective
motions are not affected by the solar rotation, but below which they are
affected.  Perhaps it will be possible to check this through numerical
simulations in future. One outcome of our model is that the poles 
of the Sun should be about 3 K hotter than the equator, another
testable prediction.

\bibliographystyle{iaulike}

\end{document}